\documentstyle{mn}

\def\ebv{E(B$-$V)}

\def\gsim{\;\lower.6ex\hbox{$\sim$}\kern-7.75pt\raise.65ex\hbox{$>$}\;}
\def\lsim{\;\lower.6ex\hbox{$\sim$}\kern-7.75pt\raise.65ex\hbox{$<$}\;}

\title[NGC 6253]{UBVRI CCD photometry of the old open cluster 
 NGC 6253\thanks{Based on  observations made at ESO telescopes, 
 La Silla, Chile}}

\author[Bragaglia et al.]{A. Bragaglia$^1$, G. Tessicini$^1$, M. Tosi$^1$, 
       G. Marconi$^2$, U. Munari$^3$\\
 $^1$ Osservatorio Astronomico di Bologna, Italy, 
      e-mail angela@astbo3.bo.astro.it, loiano@astbo3.bo.astro.it, 
      tosi@astbo3.bo.astro.it, \\
 $^2$ Osservatorio Astronomico di Monte Porzio, Italy, 
      e-mail marconi@coma.mporzio.astro.it\\
 $^3$ Osservatorio Astrofisico di Asiago, Italy,
      e-mail munari@astrpd.pd.astro.it}

\date{}

\begin{document}
\maketitle

\begin{abstract}
 We present UBVRI photometry for the old open cluster NGC 6253. From
 comparison of the observed colour-magnitude diagrams with simulations
 based on stellar evolutionary models we derive in a self consistent
 way reddening, distance, and age of the cluster:  \ebv = 0.23--0.32, 
 (m-M)$_0$ = 10.9$\pm$0.1, metallicity roughly double than solar, and age 
 $\simeq$ 3 Gyr. The cluster has a binary sequence,
 discernible even through the field contamination, suggesting that about
 1/3 of the cluster members belong to binary, or multiple, systems.
\end{abstract}


\section{Introduction}

Open clusters are a key probe of the chemical evolution and dynamics of the 
Galaxy (e.g., Janes 1979, Panagia and Tosi 1981, Friel and Janes 1993, Tosi 
1996) and the interactions between thin and thick disks (e.g., Sandage 1988), 
since they provide more reliable information on the average stellar ages 
and radial velocities at different galactic radii (e.g., Janes and 
Phelps 1995) than field stars. They are among the very few galactic
objects for which meaningful distances can be derived over a large range,
which makes them an essential tool to constrain galactic evolution
theories. 
However, to appropriately exploit the potentialities offered 
by open clusters, homogeneous and
high quality observational material, as well as accurate data treatments,
are mandatory to avoid  misleading effects (e.g, Carraro and Chiosi 1994, 
Friel 1995). 

We have undertaken a project aimed to secure photometric observations of the 
required accuracy for a well chosen sample of open clusters mapping a grid of
ages, metallicities and galactocentric distances. The goal is to expand the
number of clusters for which good quality photometric data are available,
to constrain current theories on the dynamics, structure and chemical
evolution of the Galaxy. Relevant cluster parameters are usually obtained by
fitting theoretical isochrones to the observed color-magnitude diagrams
(CMDs). We follow a different approach, namely the comparison of observed
CMDs with synthetic ones generated by numerical codes based on stellar
evolutionary tracks (Tosi et al. 1991), which we have found to be quite more
powerful in investigations both of galactic clusters (Bonifazi et al. 1990) 
and of nearby irregular galaxies (Marconi et al. 1995). 

In this paper we examine NGC 6253, an old open cluster so far scantly studied,
located toward the galactic center ($\alpha_{1950} = 16^h 55.1^m,
\delta_{1950} = -52^{\circ}38^{\prime}$; $l^{\rm II}$ = 336$^{\circ}$,
$b^{\rm II}$ = --6$^{\circ}$). Our results on Collinder 261 (age $\geq 7
\times 10^9$ yrs and solar metallicity) have been presented by Gozzoli et
al. (1996), while studies of NGC 2506 and NGC 6603 are in progress. 

In section 2 we describe the observations and data analysis; in Section 3
we present the derived CMDs involving U,B,V,R,I filters and discuss the
problem of back/foreground contamination and the presence of a conspicuous
population of binary stars. 
Section 4 will be devoted to the comparison with synthetic
CMDs and the derivation of metallicity, age, distance and reddening, and
the findings will be discussed in Section 5.

\begin{figure*}
\vspace{14cm}
\caption{Map of the observed region of NGC 6253; North is up and East left,
and the field of view is about 7$\times$7 arcmin. This map derives from our 
photometry, so it may be incomplete (e.g., there is a very bright field star
at $\sim$ 950, 270, not visible here). Also indicated are the four stars
whose pixel and absolute coordinates are given in Table 2.}
\label{fig-map}
\end{figure*}

\begin{figure}
\vspace {14cm}
\caption{Calibration relations: residuals of the fits for the various
photometric bands.}
\label{fig-err}
\end{figure}

\begin{table*}
\caption{Journal of observations. Exposure times are given is seconds.}
\begin{tabular}{lllllll}
\hline \hline 
Field   & Date           & U & B & V & R & I \\
\hline
Central & Jul 11-18, 1993 &900,900,120,40 &900,900,30 
&600,600,300, &600,120,120, &600,600,300,180,\\
        & &               &      
&60,10            & 30,10,5         &60,30,30,5 \\
North   & Jul 11-18, 1993 &     &  60 &  30 &     &  30 \\
\hline
\end{tabular}
\end{table*}

\section{Observation and data reductions}

We have observed NGC 6253 at the Danish 1.54m telescope
located in La Silla, Chile. 
The CCD mounted in direct imaging was a Tek 1000$\times$1000, with a
pixel scale of 0.377 arcsec/pixel and a total field of 6.8$\times$6.8 arcmin. 
We observed two slightly overlapping fields, one centered at the
cluster coordinates, and the other just North of it, 
intended as external field. Standard
CCD fields (PG1323-086, PG1633+099, SA110, Mark A, T Phe; Landolt 1992)
containing a total of 18 standards both blue and red were also acquired, as
was the usual set of bias,  dark exposures, and sky flats.
Exposures were taken in the Johnson-Cousins U,B,V,R,I filters, both
very short ($\sim$ 10 seconds) to avoid saturation at the bright end, 
and longer to reach deep on the main sequence.
Table 1 gives a journal of the observations and the exposure times
for all filters, while Figure~\ref{fig-map} 
shows the central field here analysed.
Table 2 lists the X, Y coordinates in pixels of four stars, and their
equatorial coordinates, for possible conversion.
The celestial coordinates were found identifying the stars in the Digitized Sky
Survey\footnote{Digitalization operated at STScI, on material from the
UK Schmidt telescope, operated by the Royal Observatory Edinburgh and the
Anglo-Australian Observatory} images distributed on CD-ROM. Precision is
about 1 arcsec, both in right ascension and declination.

\begin{table}
\begin{center}
\caption{Pixel and equatorial coordinates for the four reference stars
 indicated in Figure 1.}
\begin{tabular}{crcc}
\hline\hline
n &Star N. & X,Y (pixel)  & $\alpha, \delta$ (2000.0)\\
\hline
1 & 2008   & 817.16, 411.41 & 16 58 49.90, $-$52 42 59.05 \\
2 & 3595   & 429.94, 722.28 & 16 59 05.75, $-$52 41 05.35 \\
3 &  890   & 133.36, 191.01 & 16 59 18.04, $-$52 44 23.63 \\
4 & 2726   & 358.67, 552.43 & 16 59 08.58, $-$52 42 08.05 \\
\hline
\end{tabular}
\end{center}
\end{table}

All the reductions have been performed in the IRAF\footnote{IRAF is
distributed by National Optical Astronomy Observatories, which is operated by
the Association of Universities for Research in Astronomy Inc., under contract
to the National Science Foundation.} environment, using 
the DAOPHOT-II routines (Stetson 1987, 1992) in a standard way. 
Stars have been found in the deepest V frame, then 
every other frame has been aligned to it, and a single list of
objects was used to derive instrumental magnitudes using PSF fitting.
We then applied an aperture correction to the instrumental 
magnitudes, empirically derived from 10-20 isolated stars.

Standard stars fields were analysed using aperture photometry. The calibration
equations were derived using the extinction
coefficients for La Silla taken from the database maintained by J. Burki
(Geneva Obs.) on the ESO/La Silla archive, accessible through {\tt www}
(http://arch-http.hq.eso.org), and are the following: 
\[ {\rm U} = u + 0.104 (\pm 0.033) \cdot  (u-v) - 5.219 (\pm 0.012) \]
\[ {\rm B} = b + 0.194 (\pm 0.024) \cdot  (b-v) - 4.640 (\pm 0.010) \]
\[ {\rm V} = v + 0.016 (\pm 0.008) \cdot  (b-v) - 4.371 (\pm 0.007) \]
\[ {\rm R} = r + 0.051 (\pm 0.016) \cdot  (v-r) - 4.476 (\pm 0.009) \]
\[ {\rm I} = i - 0.021 (\pm 0.013) \cdot  (v-i) - 2.701 (\pm 0.008) \]
where {\it u, b, v, r, i} are instrumental magnitudes, while U, B, V, R, I are
the corresponding Johnson - Cousins magnitudes. 

In the central field a total of 5107 objects were detected, but we retained
only those with internal photometric error given by DAOPHOT $\sigma \le $ 
0.1 mag in all bands. We then calibrated each frame/filter and later
averaged over the same filter to obtain the mean magnitude in each band.
For the shorter exposures, only the brightest bins were considered in the
average process. 
Furthermore we visually inspected
and corrected all cases where the to-be-averaged values differed by more than
0.15 mag: in almost all cases this happened for very faint objects.

As final result, about 4200 stars have magnitudes in all the four B,V,R,I bands.
The U exposures were much shallower and only about 1450 stars have been
detected in all the 5 filters.
A table with the 5 magnitudes and positions in pixel for all objects is
available electronically from the first author. 

\begin{table}
\caption{Completeness of our measurements for the B,V,R,I filters: this is the 
average of 15 experiments. Smaller prints indicate that the errors
were higher than 0.1 mag.}
\begin{center}
\begin{tabular}{lrrrr}
\hline\hline
mag &\% B &\% V &\% R &\% I\\
\hline
12  & 100 & 100 & 100 & 100\\
13  & 100 & 100 & 100 & 100\\
14  & 100 & 100 & 100 & 100\\
15  & 100 & 100 & 100 & 100\\
16  & 100 & 100 & 100 & 100\\
17  & 100 & 100 & 100 & 100\\
18  & 100 & 100 & 100 &  91\\
19  & 100 & 100 &  92 &  76\\
20  & 100 &  89 &  71 &  63\\
20.5&  92 &  76 &  32 &  33\\
21  &  80 &  56 &  13 &  14\\
21.5&  68 &  28 &   3 &  {\tiny 5}\\
22  &  50 &   9 &  {\tiny 2} &  {\tiny 1}\\
22.5&  31 &  {\tiny 1} &     &    \\
23  & {\tiny13} &     &     &    \\
23.5& {\tiny 2} &     &     &    \\
24  & {\tiny 1} &     &     &    \\
\hline
\end{tabular}
\end{center}
\end{table}

To estimate the completeness degree of our measurements we added artificial
stars at random positions (using the DAOPHOT routine Addstar) to the deepest
frame in each filter; we added about 10\% of stars in each magnitude
bin, distributed in colours as the real ones. 
The frame was then reduced again, using exactly the same procedure and the
same parameters and PSF as before, and we counted the recovered artificial
stars for each bin. 
The experiment was repeated 15 times per filter and in Table 3 we present the
average results, except those for the U band, where completeness reflects
the much less severe crowding conditions with comparison to the other filters
and is therefore misleading.

\begin{figure*}
\vspace{17cm}
\caption{CMDs of NGC 6253: a) U $vs$ U--B; b) V $vs$ B--V; c) V $vs$ V--R;
 V $vs$ V--I. Only stars with internal errors $\le$ 0.1 mag were considered
 in the average.}
 \label{fig-cmd}
\end{figure*}

\begin{figure*}
\vspace{9cm}
\caption{Further CMDs of NGC 6253 involving the U band: a) U $vs$ U--V; 
 b) U $vs$ U--R; c) U $vs$ U--I. Note the presence of a binary sequence
 above the cluster MS.}
\label{fig-u}
\end{figure*}

\section{The colour magnitude diagrams}

In Figure~\ref{fig-cmd} and~\ref{fig-u} we present the CMDs for NGC 6253
obtained from our reductions of the central field: 
in Figure~\ref{fig-cmd}a,b,c,d we show the
``classical'' CMDs  using the 5 filters, while Figure~\ref{fig-u}a,b,c is
dedicated to the U band and the red filters. 
Even if the field stars dominate in number, the cluster main sequence (MS)
is easily located in all diagrams, and is well defined for at least 5 magnitudes
fainter than the Turn-Off (TO). The upper MS terminates with a ``hook'' and it 
shows a clear gap just below it. The TO (the bluest point of the hook)
is located at V$\simeq$14.7, B--V$\simeq$0.75. 
The TO morphology is probably complicated by the presence of a sequence of 
binary stars (see later) which converges there in the single-stars MS. 
Well defined subgiant and red giant branches
are visible, as is a red clump at V=12.7, which we interpret as the locus 
of core-He burning stars. The magnitude distance between TO and red clump is
$\delta$V=2.0.
Also, there is a well populated sample of blue stragglers.

NGC 6325 has metallicity at the very least solar (see Section 4), and
metal rich globular clusters (e.g. Ortolani, Bica \& Barbuy 1993) often show 
anomalously extended red giant branches due to large blanketing effects. This
phenomenon has been attributed to some open clusters as well, like NGC 6791
(Garnavich et al. 1994) which is very old and slightly more metal rich than
the sun, and Cr 261 (Mazur et al.
1995) which is very old but with a metallicity probably lower than solar
(Friel et al. 1995, Gozzoli et al. 1996). Despite its larger metallicity,
our data on NGC 6253 do not show such extension and turn down of the RGB, 
unless one pretends to define it through the two redder bright stars of the
CMDs in Figure~\ref{fig-cmd}. We do not consider this lack as surprising,
though, due to the relatively small number of red giants.

From the structure of the cluster CMD we can already infer that it is old, but
not older than some Gyr, because of the presence of the MS gap 
and hooked termination. We will find (see next Section) an age of about
3 Gyr.
No comparison with previous results is possible, since this is the first
study of the cluster.

\subsection{The Lower Main Sequence}

Note that the cluster MS looks increasingly less populated going 
towards fainter luminosities: there seems to be a flattening of the present
mass function for low-mass stars. This of course reflects on the luminosity
function and the comparison with synthetic diagrams (see next Section). 
This flattening is not unusual or unexpected in old
open clusters (see the review by Friel 1995) and has already been
noticed e.g., in NGC 752 and M 67 (Francic 1989, Fan et al. 1996), 
NGC 2506 (Scalo 1986), King 2 
(Aparicio et al. 1990), and as an extreme case in NGC 3680 (Nordstrom
et al. 1995). Also in NGC 2243 (Bonifazi et al. 1990), a cluster very
similar to NGC 6253, this fading of the MS is noticeable, even if not
explicitly analysed by the authors.
A likely explanation in old clusters is dynamical relaxation with ensuing
mass segregation of higher mass stars towards the central parts and evaporation
of lower mass stars (also favoured by tidal stripping due to encounters with
massive clouds). This effect has been noticed also in younger clusters,
like NGC 225 (age 10$^8$ yr, Lattanzi et al. 1991), which has
photometry, astrometry and spectroscopy of good quality,
and even in very young clusters (age $\sim$ 10$^6$ yr, e.g. NGC 2362, Wilner \&
Lada 1991), where relaxation should not have had time yet to work out its 
effects. 

\begin{figure}
\vspace{9cm}
\caption{Fore/background contamination as shown by the V $vs$ B-V CMD of the
 control field.}
\label{fig-back}
\end{figure}

\begin{figure*}
\vspace{9cm}
\caption{CMDs of NGC 6253 at different radii from the centre:  the cluster
 features, and most conspicuously the subgiant and red giant branches  and the
 red clump, are well defined in all three panels} 
\label{fig-rad}
\end{figure*}

Recently a similar fading of the MS has been found also for a globular
cluster: Paresce, De Marchi 
\& Romaniello (1995) presented HST data for NGC 6397,
where the lower MS, close to the H burning limit, is much less populated
than expected. In this case evaporation of low mass stars from the cluster
is the favoured explanation, since the relaxation time for the cluster is
small compared to its age, and the system is close to the Galactic plane, even
if a peculiar IMF is not excluded by the authors.

\subsection{Field decontamination}

The true cluster diameter is larger than that found in the literature
(5 arcmin, Lang 1992), and one has to move 8 arcmin from the cluster centre to 
find a legitimate external field. 
Figure~\ref{fig-back} presents the CMD for the external field stars, extracted 
from the Northern frame.
Given the exposure times used, we did not saturate objects in this external
field, so the lack of stars brighter than about V=14.7 is real. The field comes
from a mixed population, since it 
exhibits both a blue main sequence extending to about V=15, and an older/farther
main sequence with TO at V=19, B--V=0.9 and relative subgiant/red giant branch. 
Given the galactic location of NGC 6253 and that resulting for this old 
component by extrapolating on the cluster line of sight, we interpret it as 
the old population across and beyond the Sagittarius spiral arm.

Fore/background contamination is severe, but not irrecoverable since cluster
and field stars can be well distinguished in the CMDs 
involving the bluer bands: the
separation is especially noticeable in the U $vs$ U--V, U--R, U--I planes (see
Figure~\ref{fig-u}). Unfortunately, the U band is also that where less stars
were detected, so we decided to attempt a cluster-field segregation in the
classical V $vs$ B--V diagram; there too the separation is apparent, and
we think we can isolate {\it bona fide} field stars with reasonable confidence,
at least on the blue side of the MS. 
Results of our classification are shown in Figure~\ref{fig-sim4},
where the heavier symbols indicate probable cluster members, and the lighter
ones probable field stars. All the red giants have been assigned to the cluster,
due to their absence in the external field. 

Besides, the cluster features are well distinguishable in all panels of
Figure~\ref{fig-rad}, where we show the run of the V $vs$ B--V CMD with
different radial distances from the cluster centre (taken by visual
inspection of the map to be at pixel 400,500). The number of field stars 
appears to decrease from panel a) to c) more dramatically than that of the 
cluster members, while the subgiant and red giant branches
and even the red clump remain well delineated: we interpret this as 
further confirmation of their being cluster members.

\subsection{Binary stars}

Presence of binary stars in open clusters is not an uncommon phenomenon: most
open clusters show in fact indications of a sizeable binary population. 
Mermilliod \& Mayor (1989, 1990) during a Coravel radial velocity survey of 10
clusters as old or older than the Hyades, found a percentage of spectroscopic
binaries around 30 \% in all of them. As examples of photometric detections,
Aparicio et al. (1990), from the scatter of the main sequence, found that about
50\% of the 
stars in King 2 are binaries. Very well defined sequences are visible
in NGC 2243, where at least 30\% of the members are binary systems (Bonifazi 
et al. 1990), in M 67 (Montgomery et al. 1993 estimate $\gsim$ 38\% of 
binaries, and Fan et al. 1996 reach 50\%) or in NGC 2420 (Anthony-Twarog 
et al. 1990 derive about 50\% of binaries). 

\begin{figure}
\vspace{15cm}
\caption{Top panels: Histograms of the colour distribution in U--R relative to 
the MS ridge line for magnitudes  in the range U=17-20. The MS peak is at 0 by 
definition, and the position of the secondary peak expected in each magnitude 
bin in presence of binary systems is indicated by  an arrow. Bottom panel:
Histograms of the U-magnitude difference of all stars relative to the MS ridge
line. The arrow indicates the expected position of equal mass binaries.}
\label{fig-bin}
\end{figure}

The case of NGC 6253 is not as well defined as these; we can clearly see
(Figures~\ref{fig-cmd}, ~\ref{fig-u} and ~\ref{fig-rad}) a 
secondary sequence of stars right above the cluster MS, at a
distance of about 0.7 mag, just what we would expect in presence of a 
significant fraction of binary systems. The evidence is more convincing when 
looking at CMDs built using the U and red filters (Figure~\ref{fig-u}), 
since the separation
between the MS and field stars is larger, and the binary sequence is better
discernible. Unfortunately, the field population smears out 
the MS and binary sequence distributions. 
In order to somewhat quantify the visual impression of a secondary
sequence, we have built the histograms of the colour 
difference (e.g. in U--R, see top panels in Figure~\ref{fig-bin}) 
between each star and the MS ridge line (interpolated by eye) at different 
magnitude levels. In all panels, except the very faintest where any feature
is wiped out by the errors, a small secondary peak to the right of the MS 
one is visible. Despite its small size, this peak provides strong support
to the binary stars hypothesis, since it appears almost exactly at the colour 
(indicated by an arrow) where the peak of equal mass binaries is predicted
to occur. Notice that this does not mean that the binaries have 
equal mass components: as discussed in the pioneering work by Maeder (1974)
and most recently by Fan et al. (1996), binary components with mass ratios $q$
lower than 0.5 are indiscernible from the single stars MS, because photometric
errors are larger than their colour and magnitude difference, whereas binaries
with 0.5$\lsim q \lsim$0.9 spread out between the single star MS and the
equal mass binary ridge where the remaining binaries concentrate. In fact,
both the appearance of the CMD and the histograms derived for NGC 6253,
look very much the same as those shown by Fan et al. (1996) for M 67, where
the fraction of binaries is estimated around 50\% and their mass ratios
randomly distributed between 0 and 1.

An alternative test is to look at the distributions of magnitude differences
between stars and the MS ridge line at each colour, since it should also present
a secondary peak due to binaries (see e.g. Montgomery et al. 1993 for M~67). 
The results for our cluster are not as conclusive as theirs, given the larger 
field contamination. We have computed the magnitude differences in 
the U, U--R CMD, and their histogram is shown in the lower panel of 
Figure~\ref{fig-bin}, with an arrow indicating the predicted location of 
objects 0.75 mag brighter than the MS: there is indeed a hint of secondary 
peak in that position, but definitely not significant. 

As it is, we cannot derive a precise  figure for the binary fraction,
since we are plagued by field contamination. 
Taking into account that the number of stars in the secondary peak
of Figure~\ref{fig-bin} histograms ranges between 0.14 and 0.27 times the
total number of stars falling in both the primary and the secondary peaks,
the apparent fraction of binary systems in NGC 6253 is approximately 20\%.
Due to the various effects discussed by Maeder (1974) and Fan et al. (1996),
this presumably corresponds to a larger actual percentage of binaries.

An alternative explanation suggested by some authors for the MS skewness is
differential extinction (due e.g. to a feeble dust lane, see Lattanzi et al. 
1991 for NGC 225). In the case of NGC 6253, we reject this hypothesis, because 
the stars on the right of the MS have 
galactic locations uniformly distributed within the cluster, and in these 
conditions differential reddening should not create a second sequence 
clearly separated from the first one as in Figure~\ref{fig-bin} but rather
a general spread of the whole MS.

\section{CLUSTER PARAMETERS}

\begin{figure}
\vspace{6cm}
\caption{CMD of NGC 6253 for the objects with $\sigma \leq0.1$. Probable
cluster members are indicated by filled circles, probable field stars by
dots (see text for details).}
\label{fig-sim4}
\end{figure}

\begin{figure*}
\vspace{10cm}
\caption{Synthetic CMDs derived from the FRANEC stellar evolutionary tracks.
Panels (a) and (b) adopt Z=0.01, $\tau$=2.5 Gyr, (m-M)$_0$=10.9 and \ebv=0.43; 
panels (c) and (d) Z=0.02, $\tau$=3 Gyr, (m-M)$_0$=10.8 and \ebv=0.33.
}
\label{fig-sim3}
\end{figure*}

We have applied to NGC 6253 the approach amply described by Tosi et al. (1991),
and already employed for two other old 
open clusters (Bonifazi et al. 1990; Gozzoli
et al. 1996). Briefly, this method represents an improvement of the classical
isochrone fitting, allowing the simultaneous derivation of age, reddening and
distance modulus, with the advantage of checking the relative number of stars
in the various evolutionary phases.
The observed CMDs of the cluster are compared with synthetic
CMDs resulting from Monte Carlo simulations of a system containing the same
number of stars above the same limiting magnitude, and with the same
photometric errors and incompleteness factors in each magnitude bin as observed
in the actual cluster. For each assumed evolution model, the resulting
synthetic CMD is translated into the empirical one by finding the values of 
reddening and distance modulus providing the best agreement with the observed 
stellar distribution, and thus providing in turn metallicity, age, reddening 
and distance modulus at the same time. 
The resulting set of values, however, is not unique, because different stellar 
evolution models may lead to quite different solutions, as shown for instance
by Gozzoli et al. (1996) for Cr 261. 
For this reason, we have derived the cluster parameters 
with three different data bases of stellar evolutionary tracks, and
evaluated both the best parameter values  and the corresponding theoretical 
uncertainties.

The synthetic diagrams simulated for NGC 6253 are based on homogeneous
sets of stellar evolution models computed for several initial metallicities:
a) the tracks with classical mixing length treatment of the convective zones 
 computed by the Frascati-Teramo group (hereinafter FRANEC),
b) the tracks with overshooting from convective cores computed by the Geneva 
 group (hereinafter GENEVA),
c) the tracks with overshooting from convective cores by the Padova group 
 (hereinafter BBC).
We have not used the tracks by D'Antona et al. (1992, hereinafter CM), applied
to Cr 261, because they are available only for stellar masses smaller than
those describing the upper CMD of NGC 6253.

Within the framework of each group of stellar models, we have performed several
simulations for any reasonable combination of age, reddening and distance 
modulus, all of which have been compared with the empirical CMD and luminosity
functions of NGC 6253. We describe below only the most significant cases, 
selected on the basis of these comparisons. 

Since the synthetic CMDs obviously apply only to cluster members, we have
selected from the stars detected in the field of NGC 6253 and with photometric
error smaller than 0.1 mag in all bands but U, only those that can be taken as 
probable members (see section 3.1). This selection restricts the sample to 641 
objects out of 4200; all the others are presumably back/foreground stars. The 
supposed members are represented in Figure~\ref{fig-sim4} with the larger 
symbols, 
the other objects with the small ones. The synthetic diagrams discussed below 
therefore assume the cluster to be populated by 641 stars. 

We have shown in the previous section that this cluster is probably
populated by a significant fraction of binary systems. To further check this
point, we have performed our Monte Carlo simulations either assuming that
all the cluster members are single stars or that a varying fraction of them
are binaries. In the latter cases, for each binary system we have assumed 
a secondary/primary mass ratio randomly extracted from a flat distribution.
Colour and magnitude of each system are assigned according to the mass ratio
and following Maeder's (1974) prescriptions. This implies, for instance, that 
a system formed of equal mass stars has the same B--V as each of the two 
stars and a magnitude corresponding to double brightness (i.e. 0.75 mag 
brighter). A system with mass ratio 0.8 shows a B--V 0.04 mag redder, and a
V 0.35 mag brighter, than the primary star; systems with $q\leq$0.6 have 
in practice the same colour and magnitude of the primary star.

\subsection{Results with FRANEC stellar models}

FRANEC sets of models follow the evolution of stars between 0.6 and 1 
M$_{\odot}$ in the central hydrogen and helium burning phases and the 
evolution of stars between
1 and 9 M$_{\odot}$ up to the onset of thermal pulses on the asymptotic giant
branch. Of the available sets with different initial helium and 
metal abundances, we have used for NGC 6253 those with Y and Z equal to (0.27, 
0.01), (0.27, 0.02), (Castellani et al. 1993).\footnote{Note 
that these tracks have been computed with the LAOL Los Alamos opacities. 
According to the authors, the effect of using instead the most recent OPAL 
Livermore opacities corresponds only to assuming a slightly larger 
metallicity (Cassisi et al. 1993). Therefore, the metallicity of the FRANEC 
models mentioned above should actually be taken as about half of their 
nominal Z value.} 

Figure~\ref{fig-sim3} (c) shows one of the best synthetic diagrams obtained
with Z=0.02 for a full population of single stars. The
adopted parameters are: $\tau$=3 Gyr, E(B--V)=0.33, (m-M)$_0$=10.8. The
shapes and locations of the MS and TO predicted by the single star models 
are correct, including the hook, the gap and the relative number of stars
in the various phases. However, all the stars on the red side of the MS are 
missing. This
inconsistency can be easily overcome by including a proper fraction of
binary systems in our simulations. In fact, panel (d) shows the same CMD as
panel (c), but for a population with 30$\%$ binaries and random $q$, and 
reproduces quite well all the observed features of the MS and TO regions.
We have performed similar simulations
with other percentages of binary systems and found that lower fractions do not 
fill the MS enough, whereas higher percentages bring too many objects above 
the subgiant branch of single stars. 

\begin{figure}
\vspace{12.7cm}
\caption{Luminosity function for the observed (dots) and synthetic
(lines) stars members of NGC 6253. Top panel: LFs corresponding to the FRANEC
models. The thick line results from adopting the actual incompleteness factors 
listed in Table 1, thin lines from that modified at the faint end (see text). 
The dotted line represents the Z=0.01 case of Figure 9 (b), 
the dashed line the Z=0.02 case of Figure 9 (d).
Central panel: GENEVA models. 
The dotted line represents the Z=0.008 case of Figure 11 (b), 
the dashed line the Z=0.02 case of Figure 11 (d), and the solid 
line the Z=0.04 case of Figure 11 (f). Bottom panel: BBC models. 
The dashed line represents the Z=0.02 case of Figure 12 (b) and the 
solid line the Z=0.05 case of Figure 12 (d). 
}
\label{fig-sim5}
\end{figure}

A striking feature of all the performed simulations, for this as well as for
all the other sets of stellar models, is that, if we adopt the incompleteness
factors of Table 3, empirically derived from the CCD frames, the resulting
synthetic CMDs systematically overestimate the number of faint stars. 
This phenomenon is apparent in the luminosity functions (LF) of 
Figure~\ref{fig-sim5}, where the dots correspond to the observational data 
and the thick dashed curve to a model adopting the same parameters as above and
the empirical incompleteness. In order to predict the right number of stars in 
the fainter bins (and consequently in the bright ones as well), one has to 
artificially alter the actual incompleteness factors below V=17. 
This discrepancy is not due to an underestimate of the incompleteness (we
have performed the artificial star test 15 times obtaining always very similar
results), but to a real
decrease of the number of low mass stars, beyond any reasonable IMF.
We are inclined to interpret this decrease in terms of evaporation from the
cluster of the smallest stars, similar to what Fan et al. (1996) have found
for M 67. In all the following we then show synthetic
CMDs and LFs (thin curves) based on the modified incompleteness which allows 
to reproduce the faint end of the observational LF.
The adopted incompleteness start to deviate from the empirical one below V=17,
and becomes a factor of 5 more severe at V=21.

Finally, the model shown in Figures~\ref{fig-sim3} (c) and (d) shows a subgiant
branch more extended in colour than the actual branch of NGC 6253 and,
as a consequence, a too red RGB. Binary stars obviously
do not help in the RGB region, although the observed spread in the 
CMD distribution of the red giants (which, at this bright magnitudes, cannot 
be attributed to photometric errors) is probably due the binary stars effect. 
Due to the increased size of post-MS stars, such effect after the exhaustion
of the central hydrogen is much more complicated than the simple shift in 
magnitude and colours assumed here; however, it is apparent that it cannot 
push the predicted distribution bluewards, as would be necessary to fit the 
data. 
A lower E(B--V) cannot be invoked to solve this discrepancy, because it would 
obviously improve the fitting of the red CMD region but worsen the blue region.
We will see in the following that this discrepancy is probably
a metallicity effect. 

Figures~\ref{fig-sim3} (a) and (b) show the best synthetic diagrams, with
and without binaries, obtained with Z=0.01. The
adopted parameters are: $\tau$=2.5 Gyr, E(B--V)=0.43, (m-M)$_0$=10.9. As in
the previous case, a 30$\%$ fraction of binary stars is required to properly
reproduce the colour and magnitude extension of the MS, as well as the global
LF (see dotted line in Figure~\ref{fig-sim5}). The synthetic subgiant branch 
shows a colour extension even larger than in the solar metallicity case and 
the RGB is too red.

To verify whether a larger metallicity could help to reproduce the observed
cluster features, we have made some further simulations with the new tracks 
with Y=0.34 and Z=0.04 kindly made available by S. Cassisi and computed with 
the new OPAL opacities and a slightly different version of the FRANEC code
(Bono et al. 1996 in preparation). This set is
however not homogeneous to the other two, due to the different input physics,
and this does not allow to use it fruitfully for our purposes.

\subsection{Results with GENEVA stellar models}

The stellar evolution tracks computed by the Geneva group take into account
the possible overshooting of convective regions out of the edges defined by 
the classical mixing length theories. The models have been computed 
for  the mass range 0.8 and 120 M$_{\odot}$ 
and several initial metallicities. For NGC 6253 we have tested the cases with
initial Y and Z (0.26, 0.008), (0.30, 0.02) and (0.34, 0.04) presented by
Schaerer et al. (1993b), Schaller et al. (1992) and
Schaerer et al. (1993a), respectively. The low mass stellar models with solar
metallicity are followed up to the early asymptotic giant branch phase
(Charbonnel et al. 1996), the
others only to the tip of the RGB, therefore only for the solar composition
cases can our synthetic diagrams show the red clump corresponding to the
central helium burning of low mass stars.
The lack of stellar models for masses below 0.8 M$_{\odot}$ and for the clump
makes the synthetic CMDs not completely comparable with that of NGC 6253,
but the cluster parameters can be derived anyway.

\begin{figure*}
\vspace{10cm}
\caption{Synthetic CMDs derived from the GENEVA stellar evolutionary tracks.
Panels (a) and (b) adopt Z=0.008, $\tau$=2.5 Gyr, (m-M)$_0$=10.9 and \ebv=0.46; 
panels (c) and (d) Z=0.02, $\tau$=3 Gyr, (m-M)$_0$=10.9 and \ebv=0.36; 
panels (e) and (f) Z=0.04, $\tau$=3 Gyr, (m-M)$_0$=10.9 and \ebv=0.32. 
}
\label{fig-sim1}
\end{figure*}

\begin{figure*}
\vspace{10cm}
\caption{Synthetic CMDs derived from the BBC stellar evolutionary tracks.
Panels (a) and (b) adopt Z=0.02, $\tau$=4 Gyr, (m-M)$_0$=10.7 and \ebv=0.30; 
panels (c) and (d) Z=0.05, $\tau$=3 Gyr, (m-M)$_0$=11.0 and \ebv=0.23. 
}
\label{fig-sim2}
\end{figure*}

Stellar tracks with lower initial metallicity have bluer MS and more extended
subgiant branch. The first phenomenon implies that the corresponding synthetic 
diagrams require larger reddenings to fit the observed distribution. Combined
with the larger colour extension of the subgiant branch, this however leads
to quite red RGBs, so that, for all the performed 
simulations, the lower the initial metallicity, the redder the final RGB. 
As a consequence, models with Z=0.008 and Z=0.02 are not able to properly 
reproduce the empirical CMD of NCG 6253, whereas models with Z=0.04 are in 
agreement with the location of both the blue and the red stars. 

Figure~\ref{fig-sim1} (a) shows the best synthetic diagram obtained
with Z=0.008 for a full population of single stars. The 
adopted parameters are: $\tau$=2.5 Gyr, E(B--V)=0.46, (m-M)$_0$=10.9.
The shapes and locations in the CMD of the MS and TO predicted by the 
single star models 
are correct, but all the stars on the red side of the MS are missing.
Panel (b)  shows the corresponding CMD for a population with 30$\%$ binaries. 
It is apparent from the shown CMDs that
with this set of tracks, as well as with the other GENEVA sets, the inclusion
of binary systems places too many bright stars right after the TO, due to
the fact that the timescale of this phase is longer in the GENEVA than in the
FRANEC and BBC models. We have checked whether we could reduce the number of
binary post-TO stars by reducing the fraction of multiple systems, but we
have not reached satisfactory solutions. In fact, any decrease of the binary
fraction has much more effect on the number of MS objects than on that
of subgiants and we end up with a clear underestimate of the MS binaries
while still having too many early subgiant binaries.

The red clump is absent because these stellar tracks do not reach that phase, 
and this absence is the reason for the slight underestimate of the number of
bright stars apparent in the luminosity function (central panel in 
Figure~\ref{fig-sim5}). The lack of stellar tracks for M$<$0.8 M$_{\odot}$ is 
the cause the abrupt fall of these theoretical luminosity functions.

Figure~\ref{fig-sim1} (c)  and (d) show the best synthetic diagram obtained
with Z=0.02 for a full population of single stars and for a population with 
30$\%$ binaries, respectively. 
The adopted parameters are: $\tau$=3 Gyr, 
E(B--V)=0.36, (m-M)$_0$=10.9. For this set of tracks the core He-burning phase
has been computed and the corresponding red clump falls in the synthetic CMD at
the same brightness (V$\simeq$12.7) as the observational one, but at redder
colour ($\Delta$(B--V)$\simeq$0.05). As for the Z=0.008 case, a lower reddening
would not solve the problem, which is clearly a consequence of the intrinsic
shape of the subgiant branch in these stellar models. 
As for the Z=0.008 case, 30$\%$ binaries provide
too many bright subgiants.

Finally, Figure~\ref{fig-sim1} (e) and (f) show the best synthetic diagram 
obtained with Z=0.04 adopting: $\tau$=3 Gyr, E(B--V)=0.32, (m-M)$_0$=10.9. 
Again, as for Z=0.008, the red clump is absent from these stellar tracks. 
Aside from this point, both the synthetic CMD and LF reproduce quite well the 
various data features. The Geneva models therefore indicate that NGC 6253
has a metallicity double than solar.

\subsection{Results with BBC stellar models}

Also the stellar evolution tracks computed by the Padova group take 
overshooting from convective cores into account, although with a treatment
different from that adopted by the Geneva group. They have been computed for 
masses between 0.5 and 120 M$_{\odot}$ and for several 
initial metallicities. They reach the tip of the asymptotic giant branch or the 
ignition of the core C-O burning phase, depending on the initial stellar mass.
For NGC 6253, we have tested the sets of tracks with Y and Z equal to 
(0.28, 0.008) by Alongi et al. (1993), (0.28, 0.02) by Bressan et al. (1994) 
and (0.352, 0.05) by Fagotto et al. (1994), available at the 
Strasbourg Data center.

As already found for the Geneva models, tracks with lower metallicity show
bluer MS, larger colour extension of the subgiant branch, redder RGB. Therefore,
models with Z=0.008 do not reproduce well the data and we do not show them
here. Models with solar metallicity are not satisfactory either.
Figures~\ref{fig-sim2} (a) and (b) show the BBC synthetic CMDs with Z=0.02 in 
better agreement with the data, respectively without or with 30$\%$ binaries. 
They assume $\tau$=4 Gyr, \ebv=0.3 and (m-M)$_0$=10.7. With the binaries
included the simulation reproduces pretty well the MS and the TO
stellar distribution both in the CMD and in the LF (Figure~\ref{fig-sim5}).
The synthetic CMD shows too red colours for the giants, 
thus suggesting that NGC 6253 has a metallicity larger than solar.

Figure~\ref{fig-sim2} (c) shows the BBC synthetic CMD in better 
agreement with the data. Figure~\ref{fig-sim2} (d) shows the corresponding
CMD with 30$\%$ of binary stars. They both assume large initial metallicity 
Z=0.05, $\tau$=3 Gyr, \ebv=0.23 and (m-M)$_0$=11.0 and reproduce pretty well 
the MS and post-MS colours, luminosities and stellar distributions
(see also the LF for the case with binaries in the bottom panel of
Figure~\ref{fig-sim5}).
With the Padova models we thus find, as with the Geneva ones, that NGC 6253 is
metal rich and $\sim$3 Gyr old.

\section{Conclusions}

Open clusters are a class well suited to study the structural and evolutionary
changes in our Galaxy. In spite of all the uncertainties in dating them, and 
of the fact that even the age ranking is not totally safe when simply taken 
from literature values, they provide ages more accurate than any other disk 
objects. Ideally one would like to determine ages
consistently, fitting uniform sets of theoretical models to clusters whose
metallicities and reddenings are well constrained by observations. This is a
tough job, given the difficulties in obtaining precise enough photometry, and
can be done only for a limited sample of clusters. Also, reddenings are usually
very badly, if at all, known. Metallicities are in a better situation, since
they have been measured for many clusters through narrow-band photometry or 
spectroscopy, both with low and high resolution (Friel 1995). Our method, with 
the advantage of determining at the same time age, distance, reddening and 
metallicity, can circumvent most of these problems and 
the main conclusions that can be derived are the following.

a) In the case of NGC 6253 we find that age and distance modulus are very
tightly derived with the synthetic CMD method. In fact, the best reproduction
of the observed cluster features is obtained assuming $\tau$=3.0$\pm$0.5 Gyr 
and (m-M)$_0$=10.9$\pm$0.1, whatever the adopted class of stellar models; a 
striking agreement if we consider that different tracks can lead to rather
different ages, as we found e.g. for NGC 2243 (Bonifazi et al. 1990) and Cr 261
(Gozzoli et al. 1996).

b) The reddening derived from our analysis of NGC 6253 is rather large, 
0.23$\leq$E(B--V)$\leq$0.32, as
expected from the galactic location of the cluster, which not only is toward
the center, but falls on the Sagittarius spiral arm.
The intrinsic differences between the various sets of adopted models are
responsible for the uncertainty on the derived value. For any given metallicity,
in fact, the GENEVA tracks are intrinsically hotter, and the BBC tracks cooler,
than the others, thus implying systematically larger and smaller reddenings 
respectively.

c) A good qualitative agreement is achieved also in what concerns the cluster
metallicity. Our method cannot provide the precise chemical abundance
of the system, since it is limited at least by the restricted number of 
chemical composition cases of the available stellar tracks. However, it is 
apparent from the results presented in the previous section that NGC 6253 must 
have a metallicity roughly double than solar, since only in this case all the
predicted locations in the CMD of MS, subgiant, RGB and red clump are 
simultaneously consistent with the empirical ones.

d) The lower main sequence shows clear indication that a fraction of
low mass stars are missing, most probably due to evaporation through
dynamical relaxation and mass segregation, as found in many other cases.
No reasonable IMF could justify such an effect.

e) The cluster has a quite distinct secondary sequence, right above the 
single-star MS, attributable to binary systems. 
A sizeable population of binary stars ($\gsim$ 20 \%) 
not at all unusual for open clusters, also leads to better agreement between
the observed and the synthetic CMDs.

With this study we increase the number of well studied old open clusters,
since,
at an age of 3.0$\pm$0.5 Gyr, NGC 6253 falls in this small sample, whose
properties like galactic locations, chemical abundances, ages can be used as
reliable tracers of disk formation and evolution.

\bigskip\bigskip\noindent
ACKNOWLEDGEMENTS

It's a pleasure to thank Laura Greggio for providing the numerical code which
has been the bulk of the adopted procedure for CMD simulations, and P.
Montegriffo whose programs we used to average and clean our measurements. We
are indebted to F. Fusi Pecci for many useful suggestions. We also wish to
thank P. Battinelli and L. Ciotti for interesting discussions about binaries
and evaporations from the clusters. The FRANEC, GENEVA and BBC evolutionary
tracks were kindly made available by their authors; in particular we are
grateful to S. Cassisi and A. Chieffi for computing some tracks for us. This
research has made use of the Simbad database, operated at CDS, Strasbourg,
France.

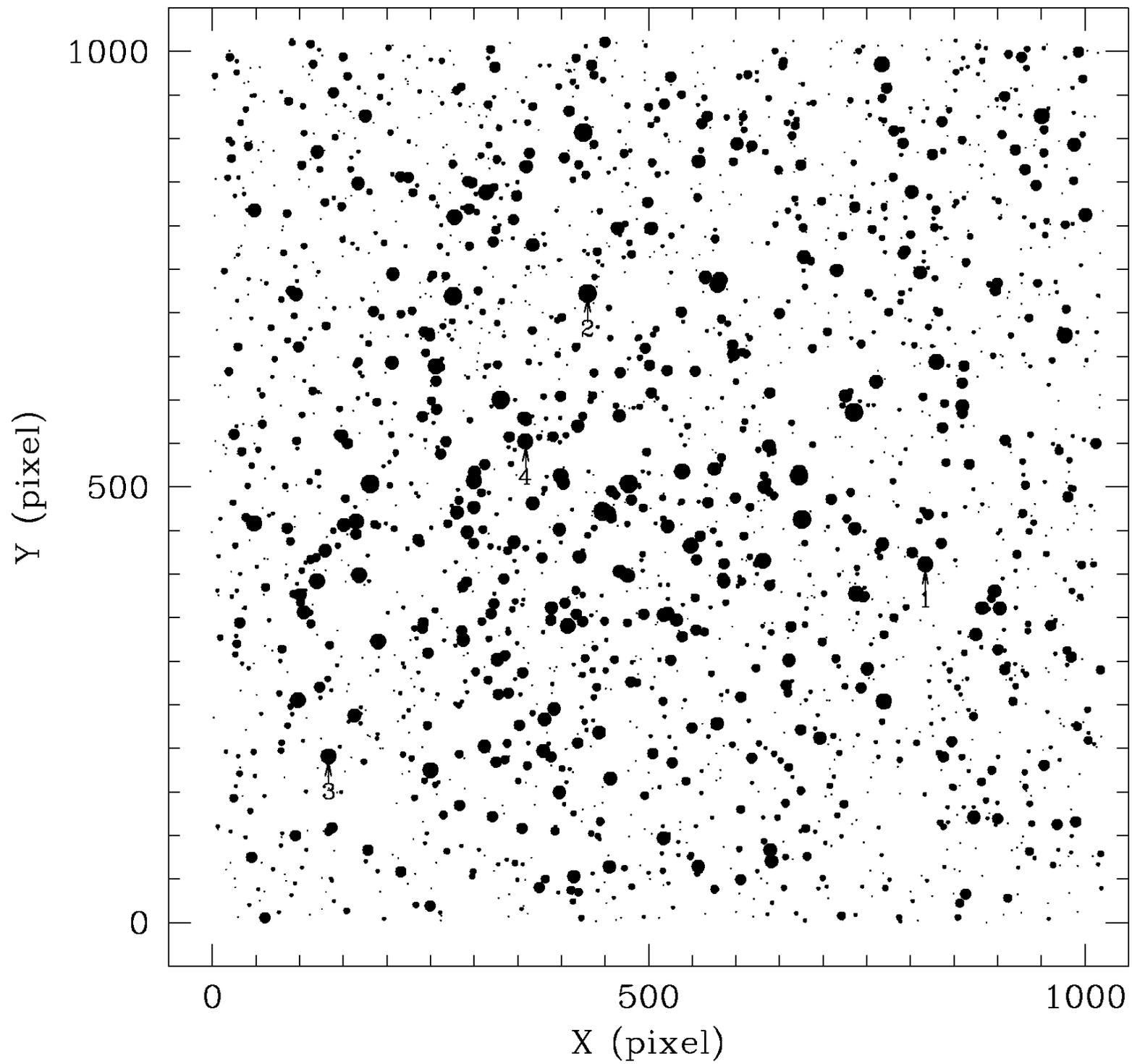
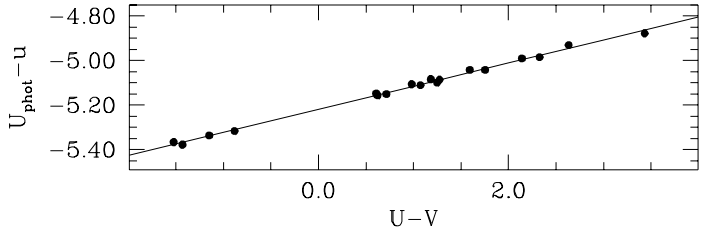
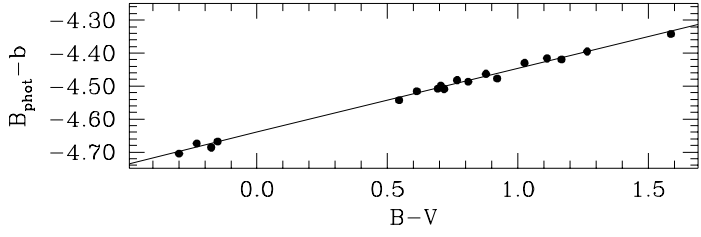
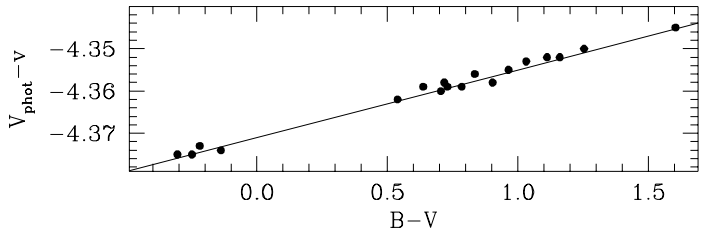
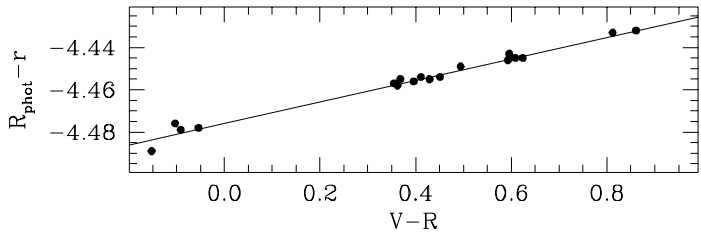
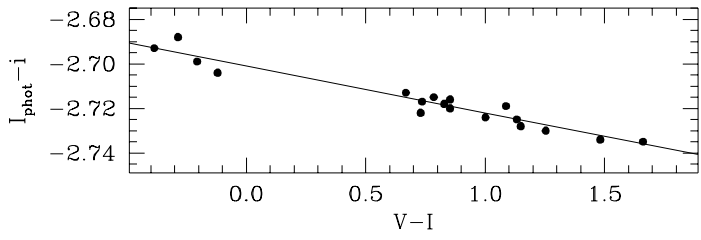
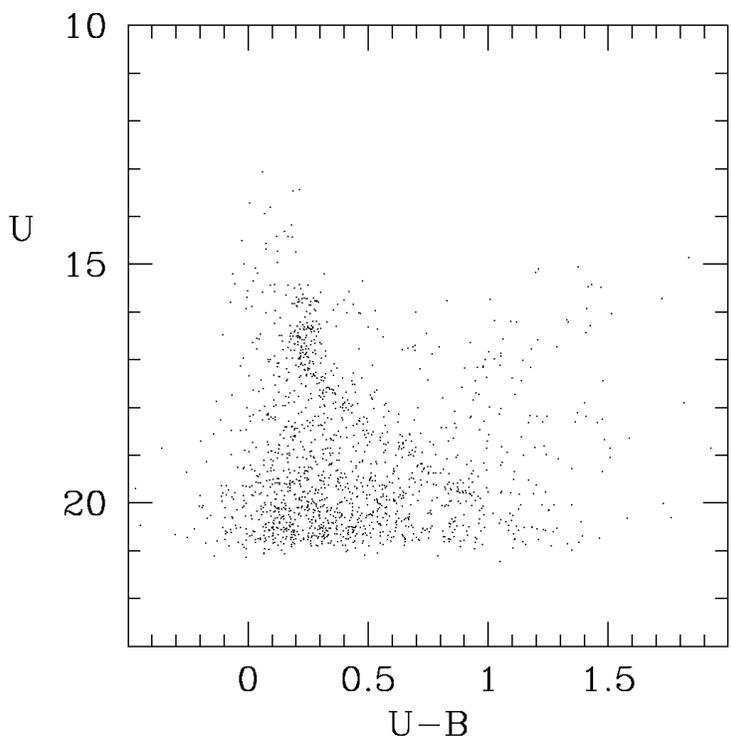
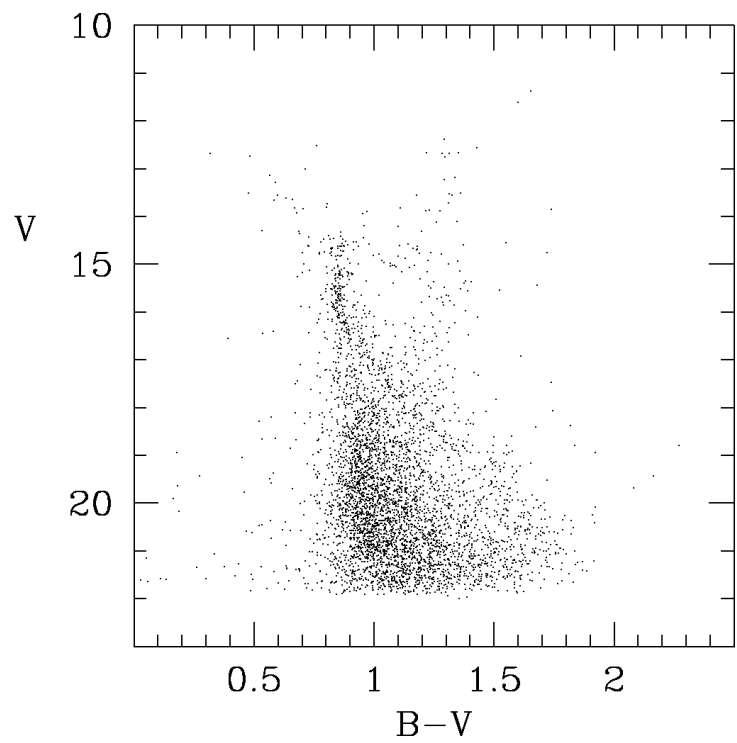
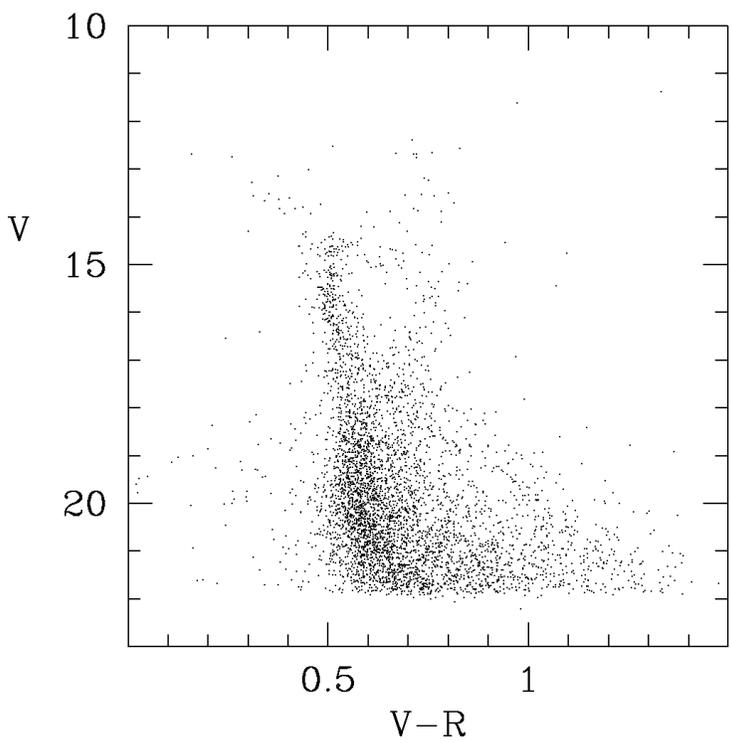
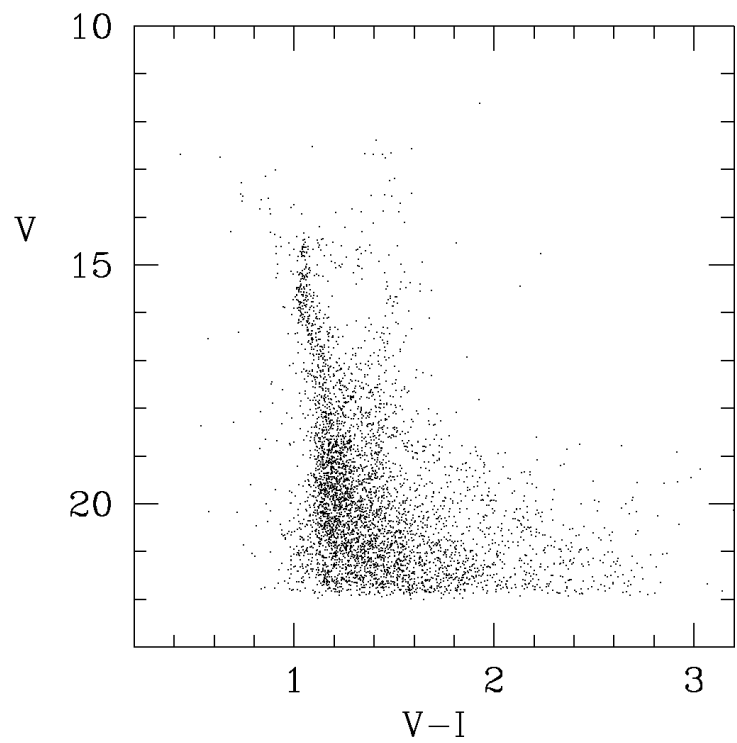
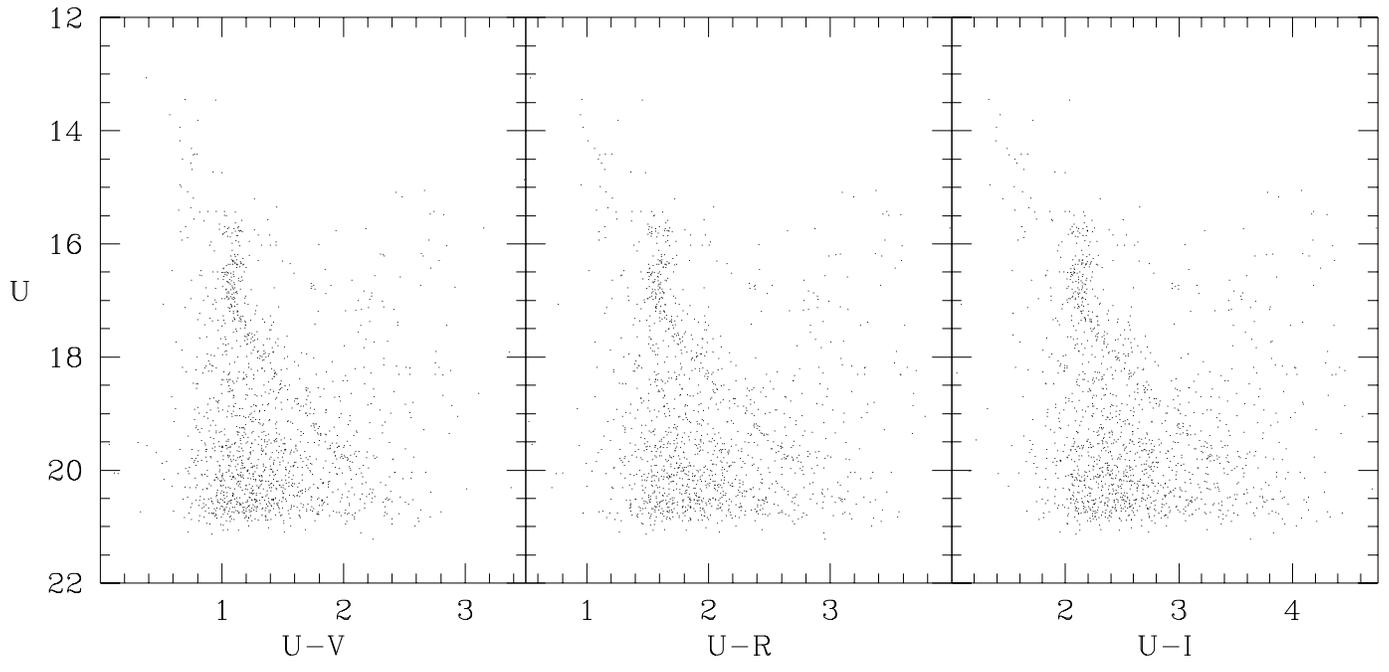
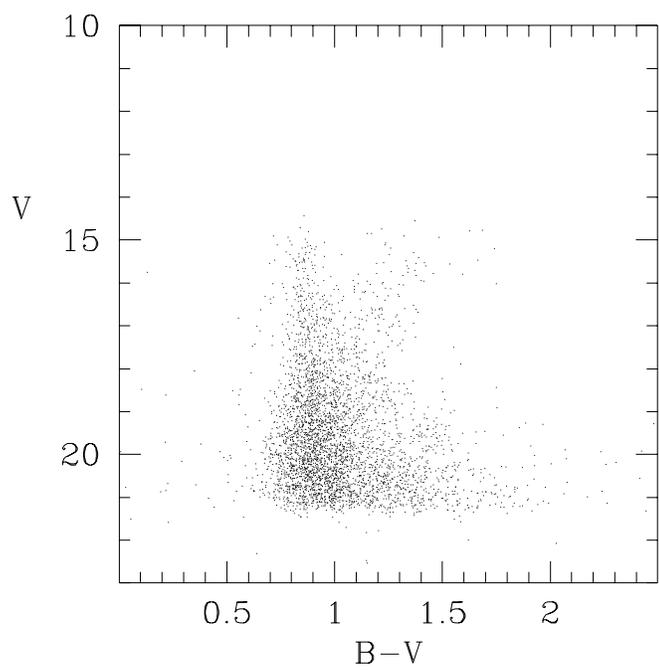
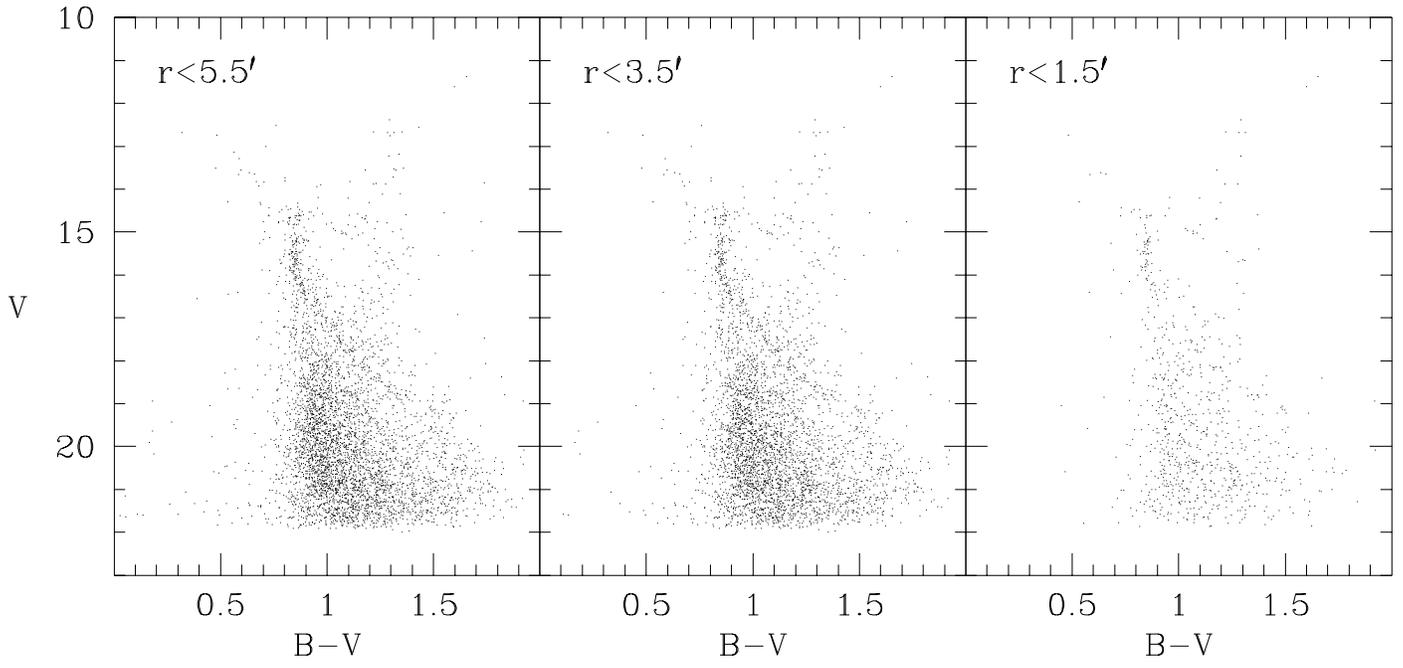
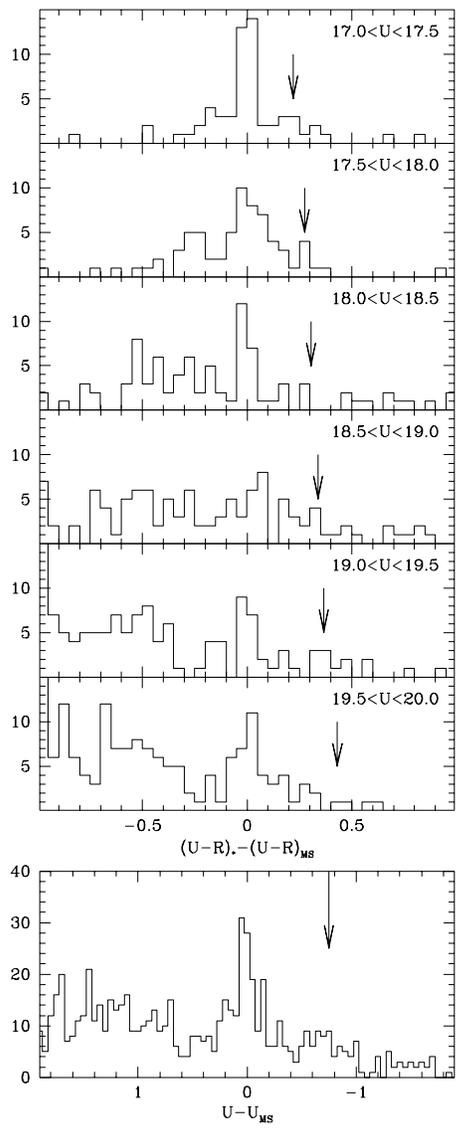
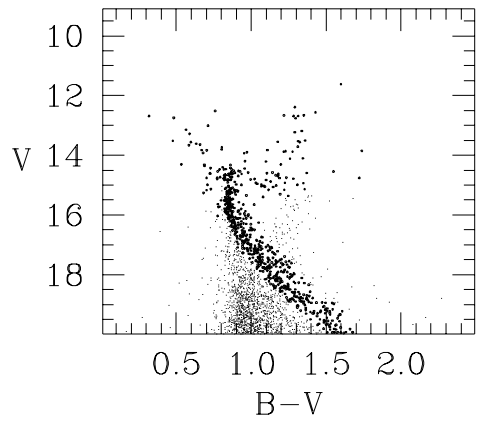
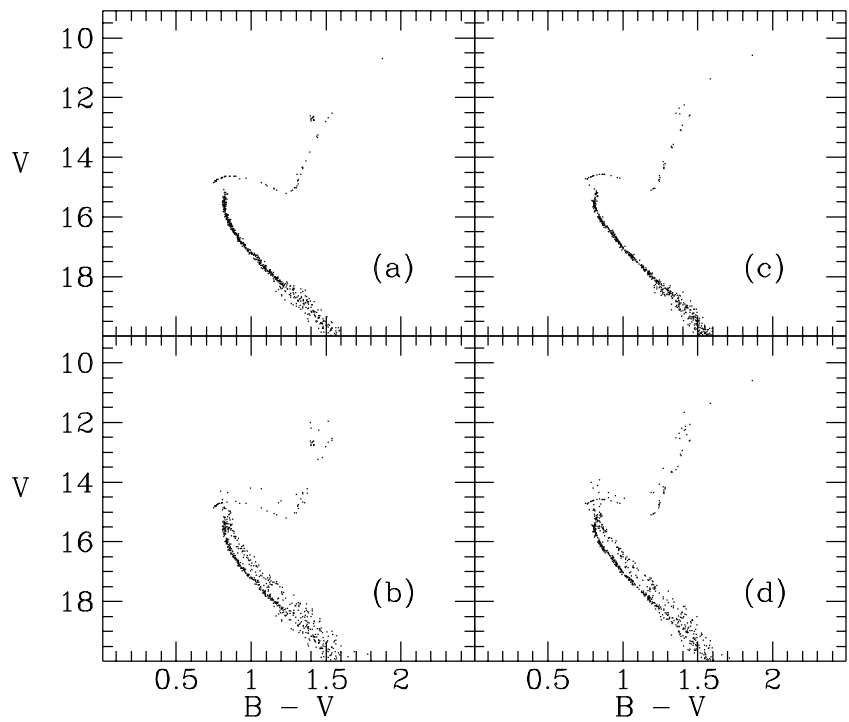
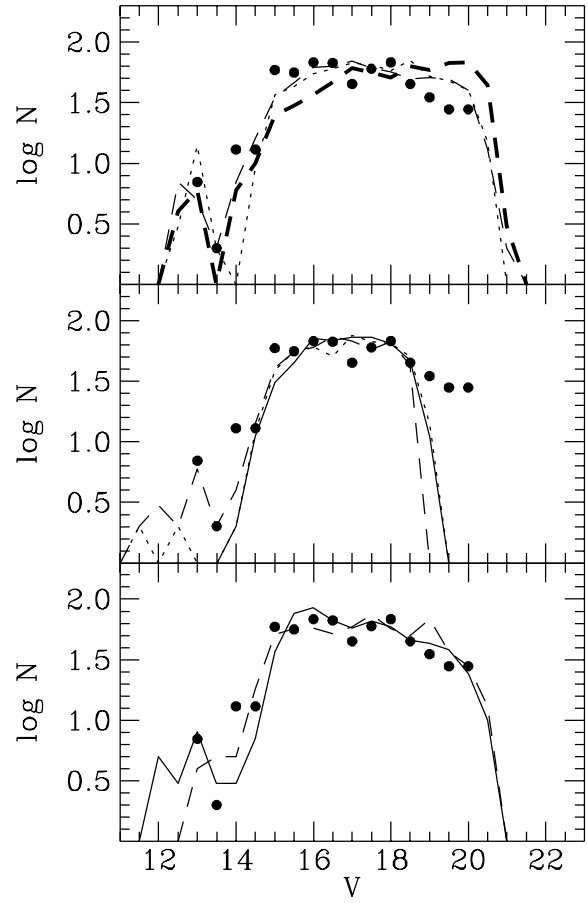
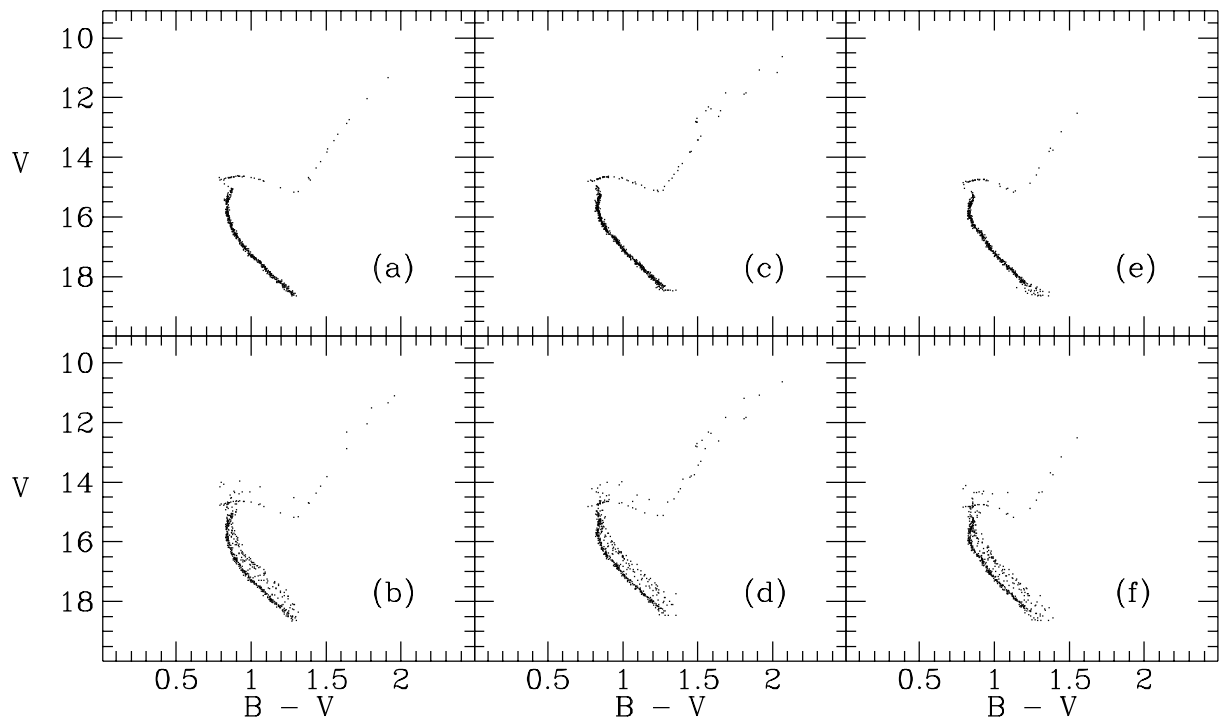
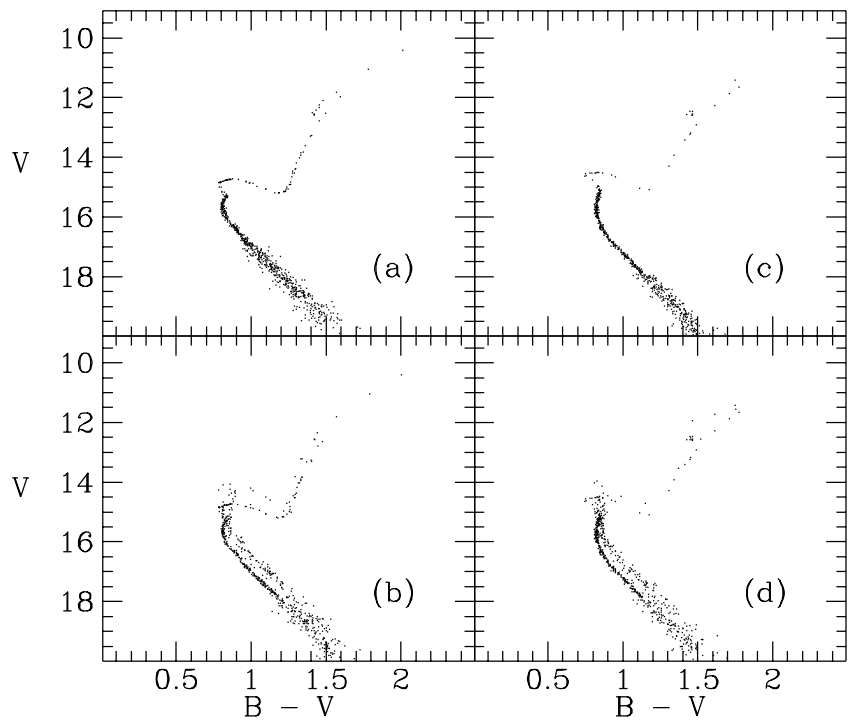
\end{document}